# Computing AC losses in stacks of high-temperature superconducting tapes

Leonid Prigozhin[1] and Vladimir Sokolovsky[2]

[1] Department of Solar Energy and Environmental Physics, Blaustein Institutes for Desert Research
Ben Gurion University of the Negev, Sede Boqer Campus 84990 Israel
[2] Physics Department, Ben Gurion University of the Negev, POB 653 Beer Sheva, Israel

E-mail:  leonid@math.bgu.ac.il,   sokolovv@bgu.ac.il

**Abstract**
Superconducting tape coils and Roebel cables are often modeled as stacks of parallel superconducting tapes carrying the same transport current. We solved, in the infinitely thin approximation, the transport current and magnetization problems for such stacks using an efficient numerical scheme based on a variational formulation of the Kim critical-state model. We also refined the anisotropic bulk approximation, introduced by Clem et al. in order to simplify AC loss estimates for densely packed stacks of many tapes; this was achieved by removing the simplifying *a priory* assumptions on the current sheet density in the subcritical zone and the shape of this zone boundary. Finally, we studied convergence of stack problem solutions to the solution of the modified bulk problem. It was shown that, due to the fast convergence to the anisotropic bulk limit, accurate AC loss estimates for stacks of hundreds of tapes can usually be obtained also using a properly rescaled model of a stack containing only ten-twenty tapes.

## 1. Introduction

Much recent interest to stacks of thin parallel superconducting strips (Fig. 1) is caused by progress in fabrication of long-length coated 2G (second generation) high-temperature superconductor tapes. The tapes can be winded into coils with a large number of turns to generate strong magnetic fields or used in Roebel cables. Characteristic feature of these configurations is the same total current in each coil turn or cable tape (in cables this is achieved by the periodic transposition of tapes, see [1,2]). Stack magnetization and transport current models are often employed in calculation of AC losses in cables. For coils, such models are also applicable if the coil inner radius is much larger than its thickness.

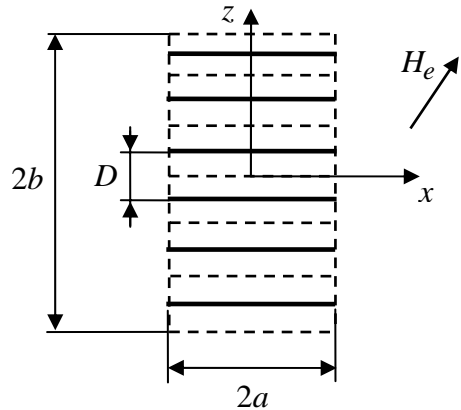

Fig. 1. A stack of tapes: *2a* and *2b* are the stack width and height, respectively; *D* is the distance between superconducting layers (solid lines). The transport current $I_{tr}$ is along the *y*-axis. The coordinate origin is in the center of the stack.

Under various simplifying assumptions, stack magnetization and transport current problems have been considered by many authors (see, e.g. [3,4,5,6]). Since the superconducting layer thickness is usually much smaller than its width, $2a$, the superconductors are often regarded as infinitely thin tapes. Indeed, for the 2G superconducting YBCO tapes the aspect (width to thickness) ratio is typically between 1000 and 10000. The distance between superconducting layers, $D$, determined by the thickness of insulator, substrate, and buffer layers, is usually between $0.01a$ and $0.2a$ [7].

Assuming the Bean critical state current-voltage relation with a field-independent sheet critical current density $J_c$ for superconducting material, Mawatari [8] and Muller [9] have found analytical solutions for infinite stacks of thin tapes in an external field $H_e(t)\bar{e}_z$ or carrying a transport current $I_{tr}(t)$ in every tape, respectively. Qualitatively, these solutions resemble those for an isolated thin strip: there are critical and subcritical current zones in every tape, in the subcritical zones the normal to tape magnetic field component is zero, although the sheet current density is nonzero. The losses occur only in the critical current zones.

Magnetization of stacks of only several tapes was simulated numerically in a number of works (see, e.g., [6,10,11]). Solving this problem for hundreds of coil turns or cable filaments is more difficult. However, for small $D/a$ ratio, solutions of the infinite stack magnetization problems are close to those for a slab of width $2a$ with the bulk critical current density $j_c = J_c/D$ and the averaged transport current per unit of the slab height, $i_{tr} = I_{tr}(t)/D$ [8,9]. It was suggested by Clem, Claassen, and Mawatari [7] that, similarly, for small $D/a$ the finite-height stack magnetization problem can be replaced by that for an anisotropic superconductor with the cross-section $[-a,a] \times [-b,b]$, the bulk critical current density $j_c$ in planes parallel to tapes, and the bulk current density $j$ satisfying $\int_{-a}^{a} j(x,z,t)dx = i_{tr}(t)$ for all $z \in (-b,b)$. For the Bean critical-state current-voltage relation, AC losses for this bulk model were estimated in [7] numerically using two additional simplifications. Namely, it was assumed that: (1) magnetic field is parallel to tape surfaces between the subcritical parts of adjacent tapes and, hence, the current density $j$ does not depend on $x$ in the subcritical zone; (2) this zone boundary can be approximated by straight lines parallel to the $z$-axis. Although these assumptions are, in simple cases, sufficiently accurate for most of the bulk superconductor, they are violated near its top and bottom surfaces and, as is noted in [7], the loss estimates become inaccurate for small $b/a$ values.

In [12,13] Weijia Yuan *et al.* solved the above magnetization problem numerically using essentially the same bulk model approximation but allowing for a magnetic field dependence of the critical current density (the Kim critical-state relation) and approximating the critical zone boundaries by two parabolas. This method was further extended to cylindrical geometry for coils with small inner radius [14].

In this work we first address the stack magnetization problem directly. We employ the variational formulation similar to that derived in [15,16] (Sec. 2) and modify the numerical scheme, based on this formulation, to deal with the transport current constraint in every tape (Sec. 3). The modified method allowed us to compute simultaneously not only the sheet current density and magnetic field but also the electric field in the tapes. Employing the critical-state models, we were able to calculate the losses in stacks containing up to a hundred of tapes using the modified numerical method. It may be noted that the variational formulation can be extended to other current-voltage relations characterizing the superconducting material [16].

The solutions obtained (Sec. 4) demonstrate peculiar current density distributions in the end (top and bottom) tapes. Although these peculiarities are typical for stack magnetization, they are lost in the simplified anisotropic bulk models [7,12,13] and, in (Sec. 5), we modify the limiting bulk model. Instead of assuming *solution properties* (subcritical current density independent of $x$) we make use of the *media property*: zero conductivity in the orthogonal to tapes direction. Similar variational formulation and numerical scheme are then derived to simulate also this anisotropic bulk superconductor; we show that our bulk model is able to describe the averaged current density also near the stack top and bottom. Furthermore, numerical solutions are obtained without any *a priory* assumption about the shape and topology of the subcritical zones, which makes our scheme universal. Comparing simulation results for the stack and bulk models, we demonstrate fast convergence, under appropriate scaling, of the stack transport current and/or magnetization problems to their anisotropic bulk limit. In fact, this convergence is so fast that, as an alternative to solving the bulk problem for estimating AC losses in stacks of many tapes, a properly rescaled stack problem with a much smaller number of tapes can be solved.

Assuming the superconducting layers are infinitely thin, we take into account only the losses caused by a magnetic flux perpendicular to their wide surface; the losses, caused by penetration of a parallel flux from the tape top and bottom surfaces, are neglected. Analysis in [7] shows that, typically, top-bottom losses are much smaller than the edge losses and can be usually neglected in applications of 2G superconductors (these losses can, however, become a significant part of AC losses in a densely packed non-inductive bifilar coil, see [17]). We also neglect currents and losses in normal metal parts of coated superconductors.

## 2. Stack magnetization problem and its variational formulation

We consider magnetization of a stack of $N$ infinitely long thin equidistant high-temperature superconducting tapes (Fig. 1),

$$\Gamma_i = \{(x, y, z_i) | -a \leq x \leq a, \quad -\infty < y < \infty\}, \; i = 1,...,N, \text{ with } z_{i+1} - z_i = D \text{ and } \sum_{i=1}^{N} z_i = 0.$$

Macroscopically, magnetization of type-II superconductors is described by the Faraday and Ampere laws supplemented by a highly nonlinear, possibly multi-valued, current-voltage relation characterizing the superconducting material. We start from the vector magnetic and scalar potential expression for the electric field in each tape,

$$\overline{E}\big|_{\Gamma_i} = -\partial_t (\overline{A}[J] + \overline{A}_e) - \overline{\nabla} V \big|_{\Gamma_i}, \tag{1}$$

where $\overline{A}[J] = A[J]\overline{e}_y$ with

$$A[J](x,z,t) = -\frac{\mu_0}{2\pi} \sum_{i=1}^{N} \int_{-a}^{a} J_i(x',t) \ln\left(\sqrt{(x-x')^2 + (z-z_i)^2}\right) dx'$$

is the vector magnetic potential (in the Coulomb gauge) of tape currents with the sheet current densities $J_i(x,t)\overline{e}_y$ and $J = (J_1,...,J_N)$, $V$ is the scalar potential, $\mu_0$ is the magnetic permeability of superconductor assumed equal to that of vacuum, and the vector magnetic potential $\overline{A}_e$ associated with the uniform external magnetic field $\overline{H}_e = H_{e,x}(t)\overline{e}_x + H_{e,z}(t)\overline{e}_z$ can be chosen as $\overline{A}_e = \mu_0(xH_{e,z} - zH_{e,x})\overline{e}_y$. The vector potentials in (1) are directed along the $y$-axis and, furthermore, $E_{i,x} = 0$ in each tape. Hence, $\partial_x V\big|_{\Gamma_i} = 0$ and the scalar potential $V\big|_{\Gamma_i}$ is independent of $x$. The $z$-component, $E_{i,z}$, is not determined uniquely (can depend on the static distribution of charges) but causes no energy losses and is, therefore, irrelevant to our consideration. Since no other variable in (1) depends on $y$, $\partial_y V\big|_{\Gamma_i}$ should be, for each tape, a time dependent constant, $C_i(t)$. Omitting index $y$ for simplicity, we can write a scalar equation for $E_i := E_{i,y}$, the only electric field component related to AC losses,

$$E_i = -\partial_t \left(A[J] + A_e\right)\big|_{\Gamma_i} - C_i. \qquad (2)$$

Implicitly, the unknowns $C_i(t)$ are determined by the conditions on the total current in each tape; we assume the transport currents are equal (and given):

$$\int_{-a}^{a} J_i(x,t)\,dx = I_{tr}(t), \qquad i = 1,\ldots,N. \qquad (3)$$

Transition from (2) to a variational current-density formulation of the magnetization problem can be carried out for a wide class of current-voltage relations characterizing superconductor material; the derivation is similar to that for a cylindrical geometry (see, e.g., [16]). Here we adopt the Kim-like critical-state model with the field dependent sheet critical current density

$$J_c(\bar{H}) = \frac{J_{c0}}{1 + \dfrac{\sqrt{k^2 H_x^2 + H_z^2}}{H_0}}, \qquad (4)$$

where $J_{c0}, k, H_0$ are constants (the model boils down to the Bean model for $H_0 = \infty$). It is assumed in the critical-state model that:

(i) the sheet current density in each tape cannot exceed the critical value,
$$|J_i(x,t)| \leq J_c\left(\bar{H}(x,z_i,t)\right);$$

(ii) the electric field (its $y$-component) $E_i$ is zero until this threshold is reached,
$$|J_i(x,t)| < J_c\left(\bar{H}(x,z_i,t)\right) \rightarrow E_i(x,t) = 0;$$

(iii) wherever $E_i$ is not zero, it has the direction of $J_i$, i.e., $J_i E_i \geq 0$.

Let the vector functions $E = (E_1(x,t),\ldots,E_N(x,t))$ and $J = (J_1(x,t),\ldots,J_N(x,t))$ denote the electric fields and sheet current densities in the tapes, respectively. Let $E$ and $J$ satisfy (2) and the critical-state model conditions (i)-(iii) with $\bar{H}[J] = \bar{H}_e + \bar{H}_J$, where $\bar{H}_J$ is the magnetic field induced by tape currents $J$. Denote by $K[J]$ the set of all functions $J' = (J'_1(x,t),\ldots,J'_N(x,t))$, $-a \leq x \leq a$, such that $|J'_i(x,t)| \leq J_c\left(\bar{H}[J](x,z_i,t)\right)$.

It is easy to see that for any test function $J' \in K[J]$ the inequality $E_i(J'_i - J_i) \leq 0$ holds at any point $(x,t)$ and for any $i = 1,\ldots,N$. Indeed, if $J_i$ is subcritical, $E_i = 0$. Otherwise, for the critical $J_i$ values, $|J'_i| \leq |J_i|$ and, by (iii), we know that $J_i E_i \geq 0$. Hence

$$\sum_{i=1}^{N} \int_{-a}^{a} E_i(J'_i - J_i)\,dx \leq 0 \qquad (5)$$

for any $J' \in K[J]$ and any $t$. The converse is also true: if $J \in K[J]$ and the inequality (5) holds for any $J' \in K[J]$, the vector functions $E, J$ obey the critical-state model conditions (ii)-(iii). Using (2), (3) and denoting by $J^0(x) = \left(J^0_1(x),\ldots,J^0_N(x)\right)$ the vector of initial current densities, we arrive at the following quasi-variational inequality formulation:

Find $C = (C_1,\ldots,C_N)$ and $J \in K[J]$ such that $J\big|_{t=0} = J^0$,

$$\sum_{i=1}^{N} \left\{\int_{-a}^{a} \left(\partial_t(A[J]+A_e)\big|_{\Gamma_i} + C_i\right)(J'_i - J_i)\,dx\right\} \geq 0 \quad \text{for any } J' \in K[J] \qquad (6)$$

and also $\int_{-a}^{a} J_i(x,t)\,dx = I_{tr}(t), \qquad i = 1,\ldots,N.$

Similar formulations of the critical-state problems have been derived in [15,16] (see also [18]) and were later interpreted as a "principle of minimal magnetic energy variation" [11]. The inequality (6) is "quasi-variational" because, in models with a field-dependent critical current density, the set of admissible solutions, $K$, depends on the unknown solution $J$ itself. This complicates theoretical study of the critical-state problems, see [19]. Nevertheless, the iterations needed to resolve this additional nonlinearity in a numerical procedure can be performed simultaneously with those for computing the unknown potentials $C_i$ and our numerical scheme is almost as efficient as for the Bean current-voltage relation with a field-independent critical current density.

The variational problem (6) contains a total current constraint for each tape and, as we found, a modification of the numerical algorithm [16] is needed to solve magnetization or transport current problems with a large number of such constraints efficiently (see the next section).

## 3. Numerical method

The finite difference approximation of (6) in time yields, at each time level $t^m = m\tau$, a stationary quasi-variational inequality

Find $C^{m-1/2}$ and $J^m \in K[J^m]$ such that for any $J' \in K[J^m]$

$$\sum_{i=1}^{N}\left\{\int_{-a}^{a}\left(\left\{A[J^m]+A_e^m - A[J^{m-1}]-A_e^{m-1}\right\}\Big|_{\Gamma_i} +\tau C_i^{m-1/2}\right)\left(J_i' - J_i^m\right)dx\right\} \geq 0 \qquad (7)$$

and also $\int_{-a}^{a} J_i^m \, dx = I_{tr}(t^m), \qquad i=1,...,N.$

This inequality is equivalent to an implicit optimization problem

$$\text{Find } C^{m-1/2} \text{ and } J^m \text{ such that}$$
$$F\left(J^m, C^{m-1/2}\right) = \min_{J \in K[J^m]} F\left(J, C^{m-1/2}\right) \qquad (8)$$
$$\text{and } \int_{-a}^{a} J_i^m dx = I_{tr}(t^m), \quad i=1,...,N,$$

where $F(J,C) = \sum_{i=1}^{N}\left\{\int_{-a}^{a}\left(\frac{1}{2}A[J]\big|_{\Gamma_i} + f_i^m + \tau C_i\right)\cdot J_i \, dx\right\}$ with $f_i^m = \left(A_e^m - A_e^{m-1} - A[J^{m-1}]\right)\big|_{\Gamma_i}$

and the minimum in (8) is sought over $J \in K[J^m]$ for a fixed $C^{m-1/2}$.

Equivalence of variational inequality (7) and constraint optimization problem (8) results from the following general observations (see [20], ch.1, par.1). If $F$ is a convex functional which achieves its minimum over a convex set $K$ at a point $J$, then $F(J) \leq F((1-\theta)J + \theta J')$ for every $J' \in K$ and any $\theta \in [0,1]$. Hence

$$\frac{d}{d\theta}F\left((1-\theta)J + \theta J'\right)\bigg|_{\theta=0} \geq 0$$

for all $J' \in K$. The converse is also true and, computing this derivative for the functional in problem (8), one arrives at inequality (7).

One can seek this problem solution, $\{J^m, C^{m-1/2}\}$, iteratively as follows:

$$J^{(n+1)} = \arg\min_{J \in K[J^{(n)}]} F_\lambda\left(J, C^{(n)}\right)$$
$$C_i^{(n+1)} = C_i^{(n)} + \kappa\left(\int_{-a}^{a} J_i^{(n+1)} dx - I_{tr}(t^m)\right), \quad i=1,...,N, \qquad (9)$$

where arg min means the point of minimum, $F_\lambda(J,C) = F(J,C) + \lambda \sum_{i=1}^{N} \left( \int_{-a}^{a} J_i \, dx - I_{tr}(t^m) \right)^2$
is an augmented functional, $n$ is the iteration number, $\kappa$ and $\lambda$ are constants, and the initial approximations can be taken from the previous time level: $J^{(0)} = J^{m-1}$, $C^{(0)} = C^{m-3/2}$. Clearly, if the iterations converge, the limit of $\{J^{(n)}, C^{(n)}\}$ is a solution to problem (8).

The iterations should be performed numerically for a spatially discretized version of (9), see the details below. We note that it is difficult to achieve convergence of these iterations using the non-augmented functional (with $\lambda = 0$) as in [16], especially if the number of tapes $N$ is large. Stable convergence has been obtained for the augmented functional $F_\lambda$.

Dividing $[-a, a]$ into $N_x$ finite elements of the length $h = 2a/N_x$, we approximate the sheet current density in each tape, $J_i(x)$, by a constant $J_{il}$ on each element $e_l$ and compute analytically (the formulas were obtained using Matematica) the double integrals

$$\tilde{G}_{il, jk} = -\frac{\mu_0}{2\pi} \int_{e_l} \int_{e_k} \ln\left(\sqrt{(x-x')^2 + (y_i - y_j)^2}\right) dx \, dx'$$

in the finite element approximation of $F(J,C)$. Although the matrix $\tilde{G}$ is full, the integrals depend only on $|i-j|$ and $|k-l|$, i.e. $\tilde{G}_{il,jk} = G_{|i-j|,|l-k|}$, and only $N \times N_x$ out of $(N \times N_x)^2$ matrix elements of $\tilde{G}$ need to be calculated and kept in computer memory. A finite element approximation of the quadratic functional $F_\lambda(J,C)$ can, at the $m$-th time level, be written as

$$F_\lambda^h = \frac{1}{2} \sum_{i,j=1}^{N} \sum_{l,k=1}^{N_x} G_{|i-j|,|l-k|} J_{il} J_{jk} + h \sum_{i=1}^{N} \sum_{l=1}^{N_x} \left(f_{il}^m + \tau C_i\right) J_{il} + \lambda \sum_{i=1}^{N} \left( h \sum_{l=1}^{N_x} J_{il} - I_{tr}(t^m) \right)^2$$

where $f_{il}^m = A_{e,il}^m - A_{e,il}^{m-1} - \sum_{j=1}^{N} \sum_{k=1}^{N_x} G_{|i-j|,|l-k|} J_{jk}^{m-1}$ is the constant approximation of $f_i^m$ on the element $e_l$.

To approximate the set $K[J^{(n)}]$, magnetic field in the $l$-th element of the $i$-th tape can be found approximately,

$$H_{x,il} = H_{e,x}(t^m) - \frac{h}{2\pi} \sum_{(j,k) \neq (i,l)} J_{jk}^{(n)} \frac{z_i - z_j}{r_{il,jk}^2}, \quad H_{z,il} = H_{e,z}(t^m) + \frac{h}{2\pi} \sum_{(j,k) \neq (l,i)} J_{jk}^{(n)} \frac{x_l - x_k}{r_{il,jk}^2},$$

where $x_k$ is the midpoint of the $k$-th element, $r_{il,jk} = \sqrt{(x_k - x_l)^2 + (z_j - z_i)^2}$. Slightly more accurate approximation, employed in our work, was based on the analytical calculation of magnetic fields generated by the finite element currents $J_{jk}^{(n)}$. The sheet critical current densities, $J_{c,il} = J_c(\overline{H}_{il})$, are then calculated according to (4) and the discretized quadratic programming problems

$$\min F_\lambda^h(J, C^{(n)})$$
$$-J_{c,il} \leq J_{il} \leq J_{c,il} \qquad (10)$$
$$i = 1, ..., N; \quad l = 1, ..., N_x$$

are solved by the method of point relaxation with projection [20] as follows. Choosing $\{J_{il}^{(n)}\}$ as an approximation to solution of (10), for each variable $J_{il}$ one finds $J_{il}'$ solving the linear equation $\partial F_\lambda^h / \partial J_{il} = 0$ while keeping all other unknowns fixed, then sets $J_{il}'' = \alpha J_{il}' + (1-\alpha) J_{il}$. Here $\alpha$ is a relaxation parameter (the convergence was faster for under-relaxation, $\alpha < 1$). The new value of $J_{il}$ is chosen then as a projection onto the admissible set,

$$J_{il} = \max(-J_{c,il}, \min(J_{c,il}, J_{il}''))\,.$$

The procedure is repeated for all variables in turn until the solution converges up to a required tolerance. The obtained solution to (10) is chosen as $J^{(n+1)}$ and the vector $C$ is updated,

$$C_i^{(n+1)} = C_i^{(n)} + \kappa \left( h \sum_{l=1}^{N_x} J_{il}^{(n+1)} - I_{tr}(t^m) \right), \quad i = 1,...,N.$$

These iterations are performed until $J^{(n)}$ and $C^{(n)}$ converge with a given accuracy. The limits, $J^m$ and $C^{m-1/2}$, respectively, are used to compute the electric field in each tape element by means of the discretized version of (2):

$$E_{il}^{m-1/2} = -\frac{1}{\tau}\left( \frac{1}{h}\sum_{j=1}^{N}\sum_{k=1}^{N_x} G_{|i-j|,|l-k|}\left[ J_{jk}^m - J_{jk}^{m-1} \right] + A_{e,il}^m - A_{e,il}^{m-1} \right) - C_i^{m-1/2}.$$

The energy loss in the $l$-th element of tape $i$ during the time interval $\left[t^{m-1}, t^m\right]$ can be estimated as

$$P_{il}^{m-1/2} = \tau h E_{il}^{m-1/2} \left( J_{il}^{m-1} + J_{il}^m \right)/2. \qquad (11)$$

The losses in each tape and in the whole stack are the sums of these losses. Losses per cycle can be determined as twice the sum of these losses through half a period, e.g., with the external field or/and transport current changing from their maximum to minimum.

We note that there are simpler ways to calculate the AC loss per cycle in a model with the Bean current-voltage relation (see, e.g., [12]). However, these methods are inexact if a more realistic Kim-like model relation is employed and may be inapplicable if a transport current and an external magnetic field are applied simultaneously; below, we use only the direct calculations of losses based on the approximation (11).

**4. Simulation results**
We present our simulation results in dimensionless form as follows. All lengths are normalized by the tape width, e.g. $x^* = x/2a$; $z^* = z/2a$; the sheet current density by the critical value in zero magnetic field, $J^* = J/J_{c0}$, and $t^* = ft$, where $f$ is the current (external field) frequency. The transport current $I_{tr}$ is normalized by $2aJ_{c0}$, and the dimensionless magnetic and electric field intensities and losses in the tape (per cycle and unit of length) are:

$$H^* = H/J_{c0}, \quad H_0^* = H_0/J_{c0}, \quad E^* = E/(2a\mu_0 f J_{c0}), \quad P^* = P/(4a^2\mu_0 J_{c0}^2) \quad (12)$$

To estimate the characteristic values of these normalizing factors, we note that coated conductors are produced by deposition of a 1-5 μm film of $YBa_2Cu_3O_{6+x}$ on 20 - 100 μm thick metallic substrate, and superconducting film is subsequently covered by protective silver and stabilizing copper layers. The width of these conductors is 5-20 mm. The total thickness of the conductor with isolation is 0.1-0.2 mm. For example, the total thickness of Superpower's SCS12050 tape is 0.15 mm. The minimum of the normalized distance $D^* = D/2a$ between the adjacent superconducting layers is ~0.01 when a coil is closely wound. In real windings, the distance can be much bigger due to a thicker isolation layer and gaps for coolant. The sheet critical current density is usually of the order of 200 A/cm at 77 K and 1000 A/cm at 20 K [12,21]. The value of $H_0$ in Eq. (4) depends on many factors (temperature, manufacturing process, etc.) and changes in a wide range. For example, for Superpower's SCS12050 tape $H_0 = 1.25 \cdot 10^6$ A/m at 20 K [12,14]; in the dimensionless form this parameter is about 10. The characteristic values of the external field amplitude are 0.1 T in cables and 1 T in motors, generators, and transformers. This corresponds to dimensionless values, respectively, of the order of 1 and 8 at 20 K and 5 and 40 at 77 K. Finally, $E_0 = 2\mu_0 a f J_{c0}$ should be of the order of $6 \cdot 10^{-2}$ V/m and $P_0 = 4\mu_0 a^2 J_{c0}^2$ is about 1 J/m at $f = 50$ Hz and 20 K and is about one order of magnitude smaller for 77 K.

The following parameters of the numerical scheme where used in all our numerical examples: $\kappa = 1.3, \alpha = 0.5$. In almost all examples $\lambda = 1$ was a good choice, however, for far separated tapes it was necessary to increase this parameter. In our computations, typically, the number of grid points, $N_x$, was between 50 and 120; the time step $\tau$ equal to 0.01 or 0.02 was sufficient to obtain the needed accuracy. Solving the same problem on crude and fine meshes, we controlled the numerical approximation error in our examples. To accelerate computations, the problem (10) at each time layer was solved with the accuracy that increased gradually with the convergence of $J^{(n)}, C^{(n)}$. All calculations were done in Matlab on a PC with the Intel DualCore 6600 2.4 GHz processor.

We begin with the Bean critical-state model and a stack of twenty tapes carrying the transport current $I_{tr} = 0.7\sin(2\pi t)$ in zero external field; the stack height is equal to its width, $2b$ =1. There are zones of critical and of subcritical currents in every tape (Fig. 2). As is expected, the electric field and normal-to-tapes component of magnetic field are zero in subcritical regions; our numerical scheme reproduces these features more accurately than the approximate approaches [7, 12].

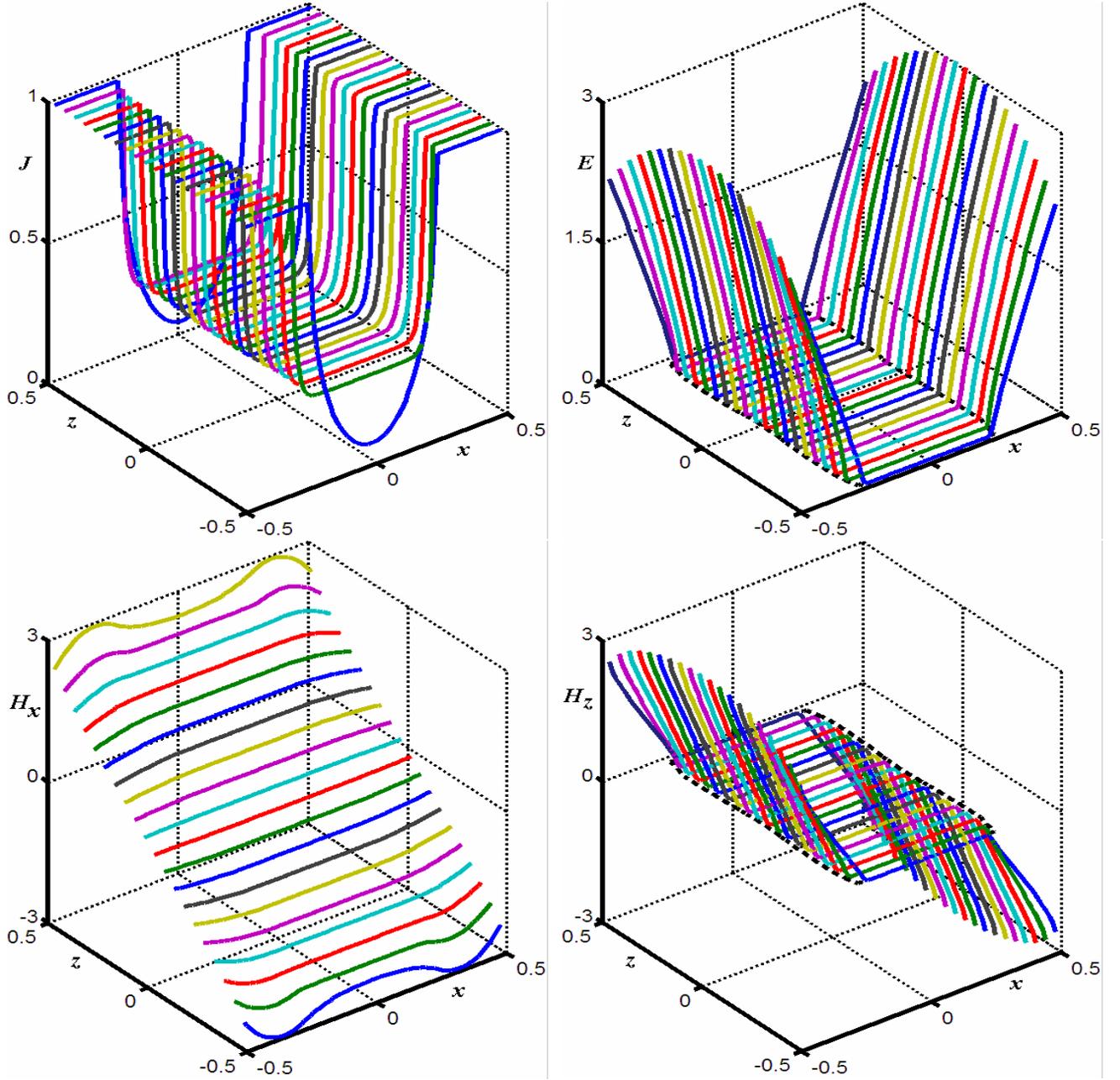

Fig. 2. A 20 tape stack: $I_{tr} = 0.7\sin(2\pi t)$, $D = 0.05$, simulation results for the Bean critical-state model. Shown at $t = 0.175$: sheet current densities (top left), electric field (top right), the *x*- and *z*- components of magnetic field in the tapes (bottom left and right, respectively). The black dash lines in the plots on the right indicate the critical current zone boundary.

We note that the parallel to a tape magnetic field component $H_x$ is discontinuous on the tape: its jump is equal to the tape current density. In the Kim model (4) and in Fig. 2 we assume $H_x(x, z_i, t) = (H_x(x, z_i - 0, t) + H_x(x, z_i + 0, t))/2$, the field in the tape "midsection".

The current distributions in the internal tapes are similar to those predicted in [7, 12], in particular, the tape current densities almost do not depend on *x* in the subcritical zones. However, in the top and bottom tapes these densities are far from uniform.

We also note that, in this example, electric field changes significantly from tape to tape: its maximum increases from 2.2 in the end tapes to 3.0 in the central ones. AC loss in a tape can, therefore, depend on the tape location inside the stack; this issue can be important for thermal state analysis and cooling system design. For a very large normalized distance between tapes, $D > 1$, the current and electric field distributions are similar in all tapes and close to those in a thin isolated strip [22]. As the distance decreases, the subcritical zone boundary admits first a )( - like shape, then changes gradually to a () - like shape (see Fig. 3, right). For $D = 0.25$ (Fig. 3, top) the penetration depth is about 20% greater in the central stack region than in the end tapes; for $D = 0.01$ (Fig. 3, bottom) the depth is, on the contrary, 70% greater in the end tapes. Loss distribution in a stack of tapes (Fig. 3, left) can be explained by the dependence of penetration depth on the inter-tape distance and tape location. We see that for $D=0.25$ the loss difference reaches 100% with the maximal loss in the most inner tapes. This result is in agreement with [14], where similar distributions have been calculated for well separated tapes using a different method. At an intermediate value, $D = 0.03$, the losses are almost uniformly distributed between the tapes; their mild variations may depend on the distance from the middle of the stack non-monotonically (Fig. 3, middle left).

Interestingly, the higher penetration depth in the end tapes can lead, since the transport current in tapes is fixed, to formation of regions of the opposite subcritical current in their central parts (Fig 4, left). These regions are observed for small inter-tape distances and intermediate (neither too small nor too big) transport currents.

Although decreasing the stack height does not eliminate the typical difference between the end and internal tape current distributions (Fig. 4, middle), the sum $\sum_{i=1}^{N} J_i$, becomes close to the known sheet current density in an isolated tape with the sheet critical current density $J_c^{\Sigma} = NJ_c$ and the transport current $I_{tr}^{\Sigma} = NI_{tr}$ (Fig. 4, right). Our simulations showed that AC losses in a thin stack are also close to losses in an equivalent isolated thin strip (see [22]). For a 10-tape stack of height 2$b$=0.05 (Fig. 4) the difference between the computed numerically AC loss per cycle and the analytically found loss in an equivalent strip is 12%; for 2$b$=0.025 and 0.0125 the differences are 7% and 4%, respectively.

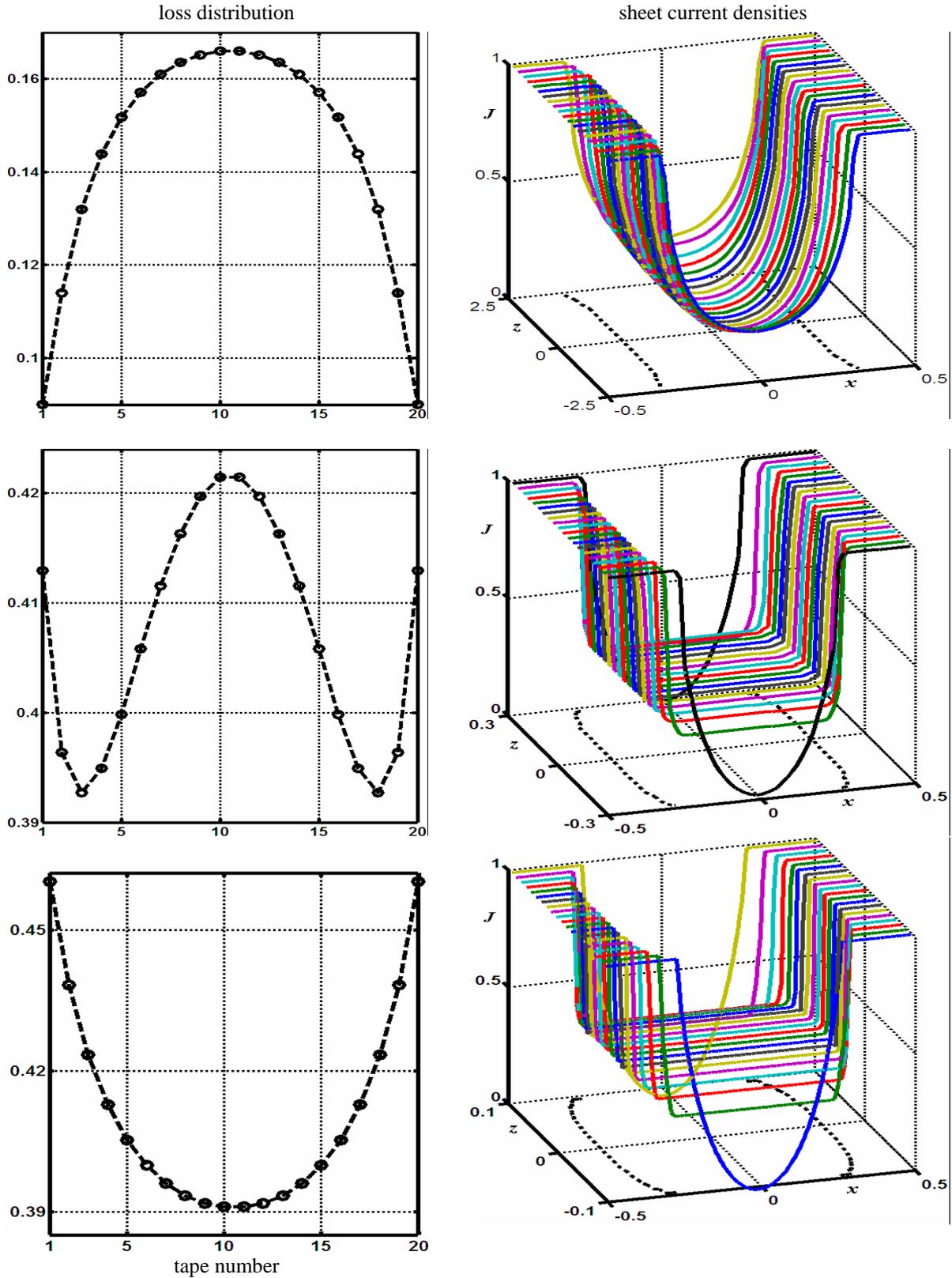

Fig. 3. Three 20-tape stacks of different heights carrying transport current $I_{tr} = 0.7\sin(2\pi t)$; the simulation results for the Bean model. Left: tape losses per period; right: sheet current densities at $t$=0.16 (black dash lines indicate the critical current zone boundary). The inter-tape distances (from top to bottom): $D$=0.25, 0.03, 0.01.

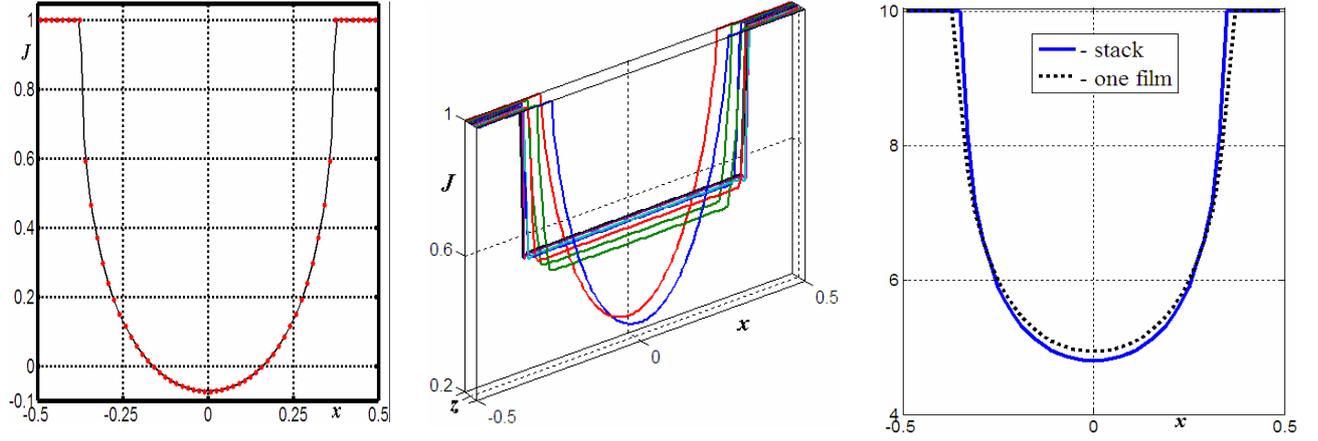

Fig 4. The transport current problem: simulation results for the Bean model.
Left: Sheet current density in the end tapes of a 20-tape stack; $D$=0.03, $I_{tr}$=0.3. Note a zone of the negative current density in the central part of the tape. Numerical solutions were obtained with $N_x = 120, \tau = 0.015$ (black line) and $N_x = 60, \tau = 0.03$ (red dots).

Middle: Sheet current densities $J_i$ in the tapes of a thin 10-tape stack; $D$=0.005, $I_{tr} = 0.7$.

Right: $\sum_{i=1}^{N} J_i$ (solid line) for the 10-tape stack and the current density in an equivalent isolated thin strip (dotted line), the analytical solution [22].

Let us now set $H_0 = 4$ in the Kim-like constitutive relation (4), which is physically reasonable, and consider the fully isotropic ($k = 1$) and fully anisotropic ($k = 0$) cases. Let us, as in our first example, consider the 20-tape stack of height $2b$=1 and apply a periodic transport current, now $I_{tr} = 0.5\sin(2\pi t)$; we assume zero external magnetic field. The computed current densities, electric and magnetic fields are shown in Fig. 5 for $t$=0.18. With high accuracy, $H_z$ and $E_i$ are again zero in the subcritical zones. The numerical scheme remains almost as efficient as for the Bean model. Thus, for $N_x = 100$ and $\tau = 0.01$, solving the problem for $0 < t < 3/4$ to estimate AC losses per period took 55 and 48 minutes of computer time for the anisotropic and isotropic cases, respectively, and the calculated total losses were $P$=4.23 and $P$=4.88. These loss estimates are very robust: using $N_x = 25$ and $\tau = 0.025$ we obtained, in less than 3 minutes in both cases, the loss estimates differing not more than 1%.

In the Bean model, the maximal transport current is $I_{tr}^M = 2aJ_{c0}$ and transition into the normal state occurs in all tapes simultaneously. For the Kim-like models the situation is different. For the anisotropic case ($k = 0$) the critical state is reached first in the middle of the stack because there $H_z$, the component of magnetic field determining the sheet critical current density $J_c$, is the highest. The modulus of magnetic field is stronger in the end tapes; hence, for the isotropic case ($k = 1$) the critical state is first reached in the top and bottom tapes. Gradually increasing the transport current in the 20-tape stack, we determined $I_{tr}^M \approx 0.68$ and $I_{tr}^M \approx 0.57$ for the anisotropic and isotropic cases, respectively. These values should be regarded as self-field critical currents for the stack.

Problems with non-zero external magnetic field can be solved in a similar way (an example is given in the next section).

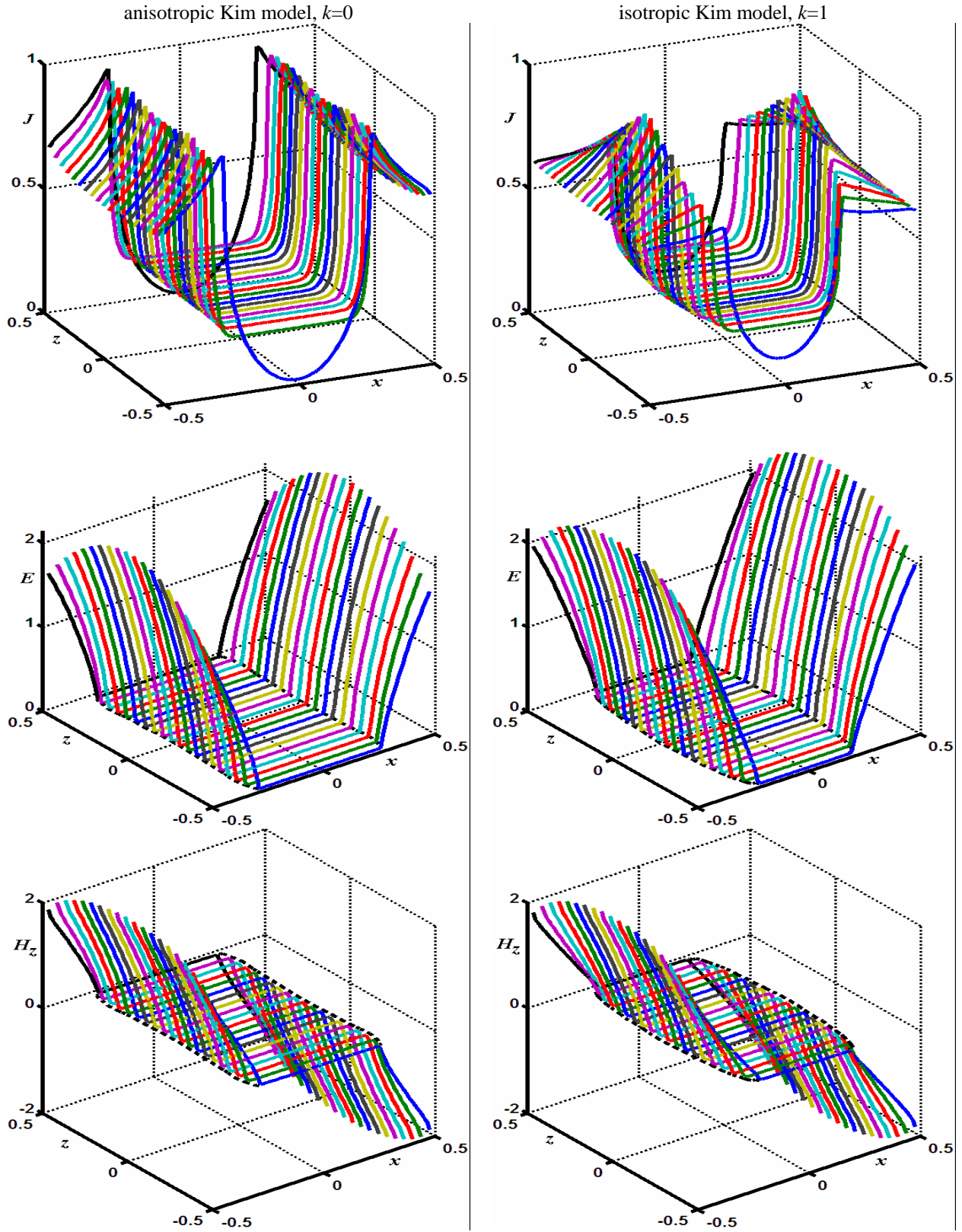

Fig. 5. Simulation results for a 20-tape stack of height $2b$=1 with the transport current $I_{tr} = 0.5\sin(2\pi t)$ in zero magnetic field. Shown for $t$=0.18: sheet current densities, electric field, and the normal magnetic field component. The Kim-like critical-state models: the anisotropic (left) and isotropic (right) field dependence of the critical current density (4). Black dash lines indicate the critical current zone boundaries.

## 5. Anisotropic bulk limit

For stacks with a large number of tapes the computations are time consuming and an anisotropic bulk superconductor model has been proposed in [7] as an approximation, accurate if $D/a$ ratio is small; this approximate model was further extended in [12,13]. As we noted above, numerical solution of the bulk magnetization problem was based in these works on two additional simplifications: (1) the current density in subcritical zone is independent of $x$; (2) the zone boundary is fitted by two straight lines or parabolas. Here we avoid any *a priory* assumption about the shape of the subcritical zones and current density.

Let us consider an infinitely long anisotropic bulk superconductor with the cross-section $[-a,a]\times[-b,b]$ in the $xz$ plane and assume the conductivity in the $z$-direction is zero. Let the current density $j(x,z,t)\bar{e}_y$ obey the Kim model $E(j)$ relations with a similar to (4) critical current density

$$j_c(\bar{H}) = \frac{j_{c0}}{1+\frac{\sqrt{k^2 H_x^2 + H_z^2}}{H_0}}. \tag{13}$$

We place the superconductor into a uniform orthogonal external magnetic field $(H_{e,x}(t), 0, H_{e,z}(t))$ and, since there is no conductivity in the $z$-direction, can postulate that $\int_{-a}^{a} j(x,z,t)dx = i_{tr}(t)$ for all $z \in (-b,b)$. The anisotropic bulk magnetization or transport current problem should be regarded as a homogenized version of the corresponding stack problem. In particular, AC losses in this bulk superconductor are expected to be similar to those in a stack of tapes with $D = b/N$, $J_{c0} = j_{c0}D$, and $I_{tr}(t) = i_{tr}(t)D$ if $D/a \ll 1$.

To investigate the anisotropic bulk model we used a variational formulation and a numerical scheme similar to those for the stack model described above. As in the stack case, $E_x = 0$, $E_y$ is determined by the Faraday and Ampere laws supplemented by current-voltage relations of the critical-state model, and $E_z$, which remains undetermined, gives no contribution to losses because the current in the $z$-direction is zero. Since both the variational formulation and numerical scheme are similar, we mention only the changes caused by the transition to the bulk. Setting $E := E_y$ to simplify the notations, we obtain, similarly to (2),

$$E = -\partial_t \left(A[j] + A_e\right) - C \tag{14}$$

but now $C = C(z,t)$. Let the pair $(j,E)$ satisfy the critical-state model conditions with the critical current density $j_c[\bar{H}](x,z,t)$ determined by (13) ; here the magnetic field $\bar{H} = \bar{H}[j]$ is the sum of the field induced by the superconductor current with density $j$ and the external field. Denoting by $K_0[j]$ the set of functions $j'(x,z,t)$ satisfying $|j'| \leq j_c[\bar{H}]$ for all $x,z,t$, we arrive at a similar to (7) variational problem,

Find $C=C(z,t)$ and $j \in K_0[j]$ such that $j|_{t=0} = j^0$,

$$\int_{-b}^{b}\int_{-a}^{a} \left(\partial_t \left(A[j] + A_e\right) + C\right)\left(j' - j\right) dx\,dz \geq 0 \quad \text{for any } j' \in K_0[j] \tag{15}$$

and also $\int_{-a}^{a} j(x,z,t)dx = i_{tr}(t)$ for all $z \in [-b,b]$.

We discretize the quasi-variational inequality (15) in time and use a uniform $N_x \times N_z$ rectangular finite element mesh in space with piecewise constant approximation of functions $j$ and $C$ on each time level. For simplicity, we used square finite elements (which slightly restricted possible $b/a$ ratios) and found, as in [15], the double surface integrals

$$G_{|i_1-i_2|,|j_1-j_2|} = -\frac{\mu_0}{2\pi}\int_{e_{i_1 j_1}}\int_{e_{i_2 j_2}} \ln\left(\sqrt{(x-x')^2+(z-z')^2}\right)dxdx'dzdz'$$

analytically for coinciding and adjacent finite elements $e_{i_1 j_1}$ and $e_{i_2 j_2}$. For other element pairs the integrals are not singular and were calculated numerically using an efficient sparse cubature formula of seventh order [23]. Computing this matrix is fast (needs less than a second for all meshes we used). The iterative numerical algorithm for solution of (15) was similar to that for the stack problem described above; in particular, the usage of augmented functional was necessary also for the bulk problem.

The anisotropic bulk model is an approximation of the stack model and we will now try to evaluate its accuracy and applicability. Let us start with the stack of height $2b = 2a$ considered above (Fig. 5, left) containing only 20 tapes; the same scaling of variables is used, $k=0$, $H_0=4$ (the anisotropic Kim model). Linearly increasing the transport current, $I_{tr}(t)=t$, we obtain (see Fig. 6, left) that the sheet current density in the tapes, $J$, is indeed close to the rescaled bulk density $jD$; the electric fields calculated using the two models are also very close (Fig. 6, right).

For the same 20-tape stack carrying periodic transport current $I_{tr}(t)=0.5\sin(2\pi t)$ the stack model prediction for AC losses per period is 0.7% higher than the corresponding anisotropic bulk model estimate (for Kim's models with both $k=0$ and $k=1$). For the stack of the same height containing only 10 tapes, the difference between the stack and bulk model predictions of AC losses is about 4%; the difference is about 1.5% and 0.5% for the 10-tape stacks of height 0.5 and 0.25, respectively (for both values of $k$).

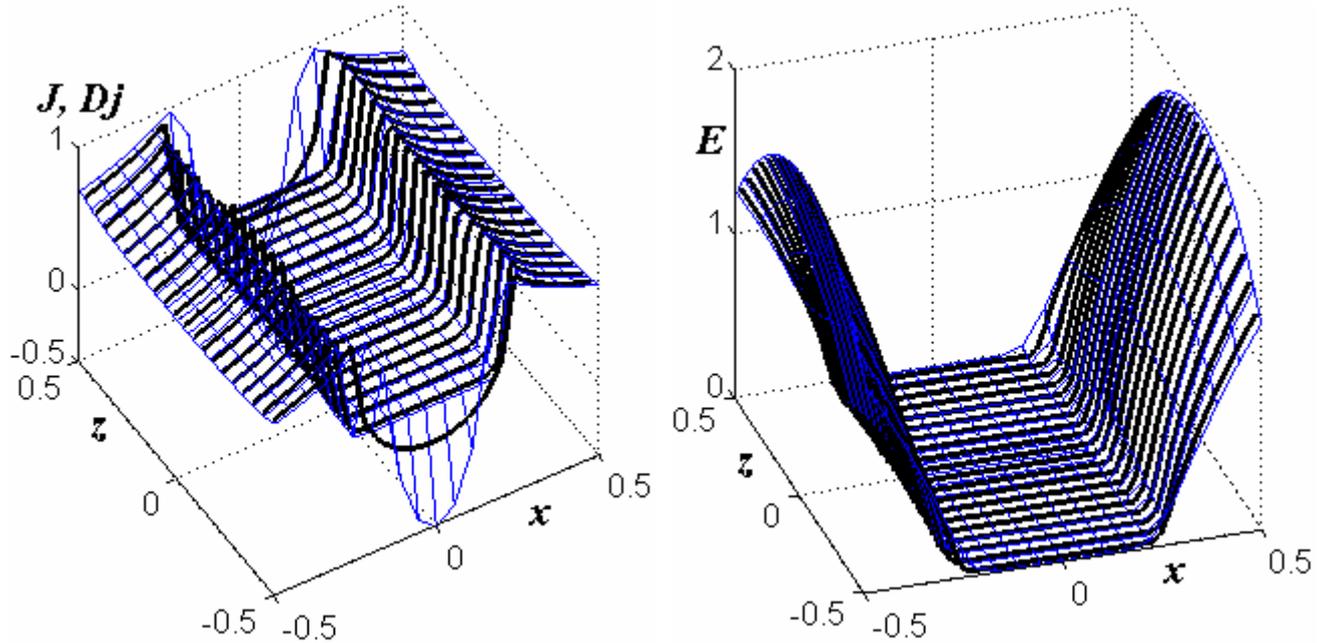

Fig. 6. Comparison of solutions for the 20-tape stack and the anisotropic bulk superconductor. Growing transport current, $I_{tr}=t$; the Kim model ($H_0=4$, $k=0$). Shown for $t=0.5$: left - the tape sheet current density $J$ and the rescaled bulk density $Dj$; right – the electric field $E$. Thick black lines – solution for the stack model; blue lines – the bulk model solution.

As another example, we compare magnetization of a 10-tape stack of unit height ($D = 0.1$) and a similarly scaled anisotropic bulk superconductor, now in a growing uniform external magnetic field parallel to the $z$-axis, $H_{e,z} = t$ (Fig. 7 corresponds to $H_{e,z} = 1.7$). Even though the number of tapes is small, electric field in the tapes is very close to the electric field in the bulk superconductor. The sheet current densities in tapes, $J_i$, are also close to the corresponding rescaled bulk density, $jD$, especially in the critical current region, where the losses occur. According to our computations, for $H_{e,z}(t) = 2.5\sin(2\pi t)$ the losses per period were 5.74 and 5.95 for the stack and bulk models, respectively; the difference is 3.5%.

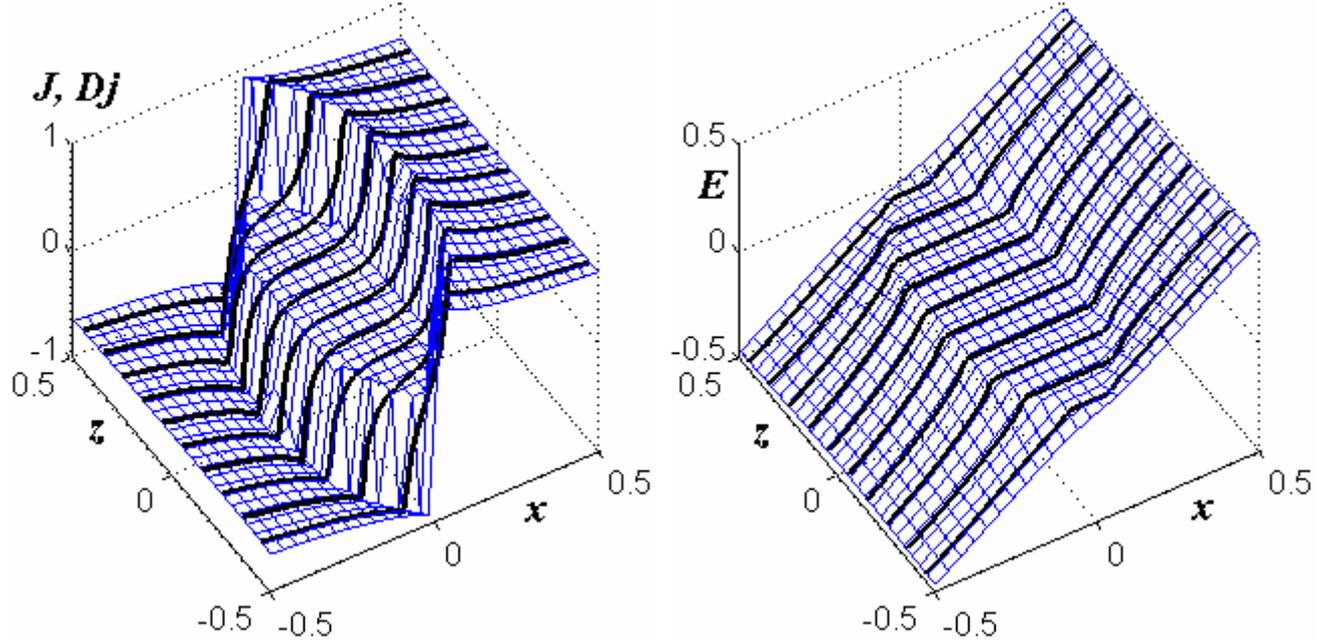

Fig. 7. Comparison of solutions for the 10-tape stack and the anisotropic bulk models. Growing external field $H_{e,z}(t) = t$, the Kim model ($H_0 = 4$, $k = 1$). Shown for $t = 1.7$: tape sheet current densities $J$ and rescaled bulk current density $Dj$ (left); the electric field $E$ (right). Thick black lines – solution for the stack model; thin blue lines – the bulk model solution.

As the transport current or external magnetic field grows, the subcritical zone shrinks. It is interesting to compare the moving boundary of this zone for stack and bulk models, scaled appropriately. Although stacks in two previous examples (Fig. 6 and 7), contain a small number, 20 and 10 tapes, respectively, the subcritical zones in the two models are very close, see Fig. 8.

Let us discuss the magnetization example in more details. Since we assume the tapes are infinitely thin, only variations of the normal to tapes field component, $H_{e,z}(t)$, can induce shielding currents and cause losses. The parallel magnetic field component, $H_{e,x}(t)$, may influence only the critical current density (4) in the Kim model. In the absence of transport current, the sheet current densities in tapes are odd functions; in every tape the condition $\int J_i(x,t)dx = 0$ holds with $C_i(t) = 0$ (for our choice of the potential $\bar{A}_e$). The situation is similar for the anisotropic bulk model where, if $i_{tr} = 0$, the condition $\int j(x,z,t)dx = 0$ holds with $C(z,t) = 0$. Note that for $H_{e,x} = 0$ and zero transport current the distribution of current in a usual isotropic bulk superconductor model should also satisfy this condition due to symmetry. Hence, the anisotropic bulk superconductor model yields in this case the same distribution of current density as the isotropic one. The latter problem was studied before (see, e.g., [15,24]). It is known that, while the external field is not too strong, the zero-field zone touches the superconductor cross-section boundary at two points, $(0,\pm b)$; it detaches from the boundary completely as the field grows further. This typical behavior (see Fig. 8, right), is not well described by the solutions [7,12], less accurate near the stack top and bottom and inconvenient, as all front-tracking methods, if the free boundary topology changes. Note that an adjustment of the numerical procedure [12] is needed also if the external field and/or transport current are non-monotone functions of time (see [13]) because new critical zones appear. Our scheme is of the free-marching type: it computes the current density, the critical current zones are then determined as the regions where the density is critical, and no *a priori* assumption about these zones topology is needed.

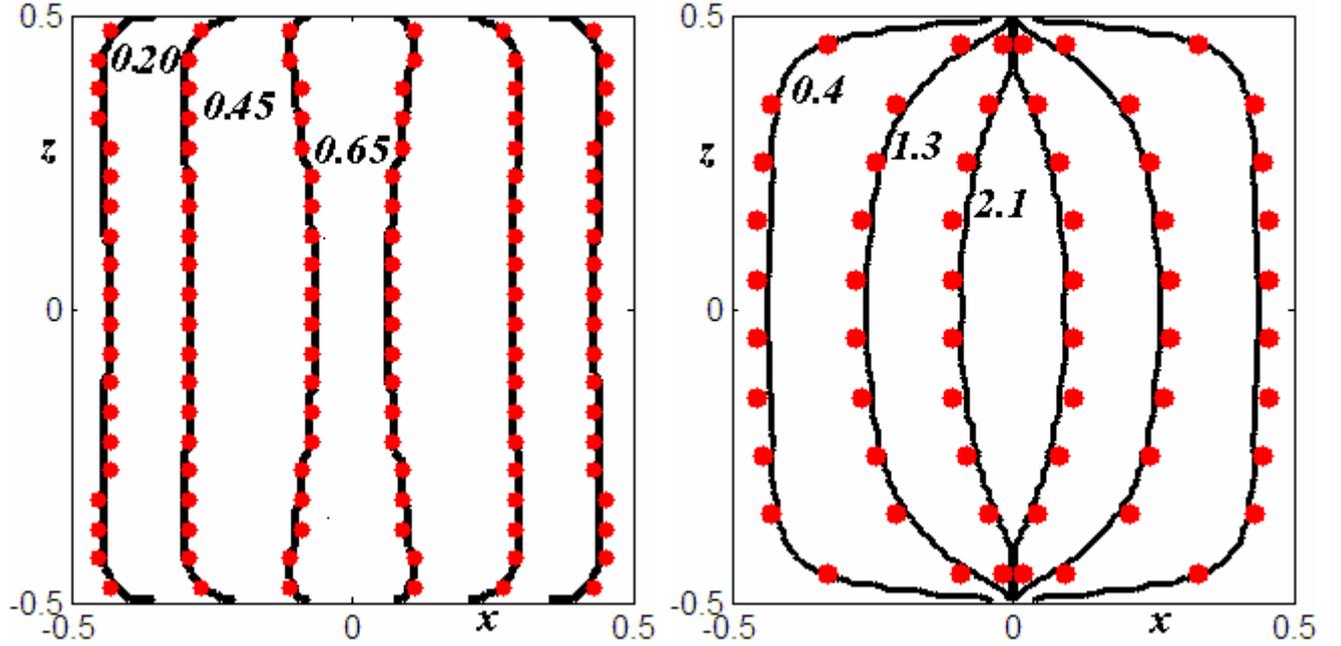

Fig. 8. The moving subcritical zone boundary: the stack model (red dots) and the anisotropic bulk superconductor model (black lines). Left: a 20-tape stack (as in Fig. 6) carrying growing transport current; $I_{tr} = 0.20, 0.45, 0.65$. Right: a 10-tape stack (as in Fig. 7) in a growing external field; $H_{e,z} = 0.4, 1.3, 2.1$.

To study further the convergence of stack model solutions to the solution of the bulk model, we fixed the bulk problem by replacing $J_{c0}$ by $2aj_{c0}$ in transition to dimensionless variables (see sec. 4). The anisotropic bulk model was used then to compute AC losses for a variety of periodic external fields $H_{e,z}(t)$ and/or transport current densities $i_{tr}(t)$, several $b/a$ ratios, and both types (the isotropic and anisotropic) of the Kim-like critical-state model relations. For each case we also computed the losses in stacks with $N$=5, 10, 20, and 40 tapes, setting $J_{c0} = j_{c0}D$, $I_{tr} = i_{tr}D$, where $D = b/N$. The accuracy of numerical solutions was controlled by solving the same problem on different meshes. In all cases the AC losses in the stack model converged quickly to those in the bulk model. Typically, for $D/a \sim 0.05$ the difference did not exceed 2% and was often much smaller.

Accurate numerical solution of the bulk problem can take more time than solving the corresponding stack problem with a few tens of tapes. Hence, an alternative approach is to replace the stack problem with many tapes by a series of appropriately scaled simpler problems for stacks of the same height, starting, e.g., with the inter-tape distance-to-width ratio, $D/2a$, equal to 0.1 and gradually increasing the number of tapes until the convergence of AC losses is reached. Usually, the convergence is fast. As an example, we calculated AC losses in a 100-tape stack of height 2 assuming the Bean model and periodic transport current $I_{tr} = 0.7\sin(2\pi t)$. Direct solution of this problem on the time interval (0, 0.75) with $N_x = 40$ and $\tau = 0.025$ took about 18 hours of CPU time and yielded the value $P = 142.1$ for the losses per period. Scaled problems with a smaller number of tapes allowed us to obtain accurate loss estimates much faster (see Table 1). The same mesh and time step were used.

| Number of tapes | CPU time | AC loss estimates, accuracy in % |
|---|---|---|
| 10 | 1.5 min | 8.5 |
| 20 | 6.5 min | 2.6 |
| 30 | 30 min | 1.3 |

Table 1. Accuracy of AC loss estimates using scaled problems.

## 6. Conclusion

The transport current and magnetization problems for finite stacks of 2G superconductor tapes were studied, in the infinitely thin approximation, by means of numerical simulation. Our algorithm, based on the variational formulation of critical-state problems [16], efficiently solves problems with several tens of tapes in a stack. If the number of tapes reaches the order of hundred, computations become time consuming.

In such a case it is better to use a homogenized model [7], the model of an anisotropic bulk superconductor, which gives accurate AC loss estimates for densely packed stacks of many tapes. We improved the mathematical formulation of this limiting bulk problem by removing unnecessary simplifying assumptions on the current density in the subcritical zone and the shape of this zones boundary, and derived a numerical algorithm, also based on the variational formulation [16], more accurate and universal than the previous schemes [7,12,13].

Knowledge of the limiting behavior significantly simplifies modeling and estimating AC losses in problems with a large number of tapes. However, the applicability and accuracy of the bulk approximation have not been fully investigated yet. Although rigorous proof of convergence to this bulk limit remains an interesting open mathematical problem, our numerical simulations confirmed fast convergence of AC losses in stacks to those in the bulk superconductor; this is, possibly, the first direct and comprehensive check of the limit hypothesis [7]. Even for ten-twenty tapes in a stack, very good agreement of calculated losses, and also currents, electric and magnetic fields, to those obtained in the anisotropic bulk model simulations has been observed.

Another approach, proposed in this work and based on solution of appropriately scaled problems for stacks with a few tens of tapes, can be even more efficient.